\documentclass[12pt,titlepage]{utarticle}

\newcommand{\tr}{{\rm tr}}
\DeclareMathSymbol{\square}{\mathord}{AMSa}{"03}
\DeclareMathSymbol{\rightsquigarrow}{\mathrel}{AMSa}{"20}

\batchmode
  \font\bbbfont=msbm10
\errorstopmode
\newif\ifamsf\amsftrue
\ifx\bbbfont\nullfont
  \amsffalse
\fi
\ifamsf
\def\IR{\hbox{\bbbfont R}}
\def\FI{Fayet-Iliopoulos}
\def\IC{\hbox{\bbbfont C}}

\def\IZ{\hbox{\bbbfont Z}}
\def\IF{\hbox{\bbbfont F}}
\def\IP{\hbox{\bbbfont P}}
\else
\def\IR{\relax{\rm I\kern-.18em R}}
\def\IZ{\relax\ifmmode\hbox{Z\kern-.4em Z}\else{Z\kern-.4em Z}\fi}
\def\IF{\relax{\rm I\kern-.18em F}}
\def\IP{\relax{\rm I\kern-.18em P}}
\fi

\def\G{{SO(32)}}

\newdimen\tableauside\tableauside=1.0ex
\newdimen\tableaurule\tableaurule=0.4pt
\newdimen\tableaustep
\def\phantomhrule#1{\hbox{\vbox to0pt{\hrule height\tableaurule width#1\vss}}}
\def\half{\frac{1}{2}}
\def\drawbox#1#2{\hrule height#2pt
        \hbox{\vrule width#2pt height#1pt \kern#1pt
              \vrule width#2pt}
              \hrule height#2pt}

\def\phantomvrule#1{\vbox{\hbox to0pt{\vrule width\tableaurule height#1\hss}}}
\def\ZZ{\relax{\sf Z\kern-.4em Z}}
\def\sqr{\vbox{%
  \phantomhrule\tableaustep
  \hbox{\phantomvrule\tableaustep\kern\tableaustep\phantomvrule\tableaustep}%
  \hbox{\vbox{\phantomhrule\tableauside}\kern-\tableaurule}}}
\def\squares#1{\hbox{\count0=#1\noindent\loop\sqr
  \advance\count0 by-1 \ifnum\count0>0\repeat}}
\def\tableau#1{\vcenter{\offinterlineskip
  \tableaustep=\tableauside\advance\tableaustep by-\tableaurule
  \kern\normallineskip\hbox
    {\kern\normallineskip\vbox
      {\gettableau#1 0 }%
     \kern\normallineskip\kern\tableaurule}%
  \kern\normallineskip\kern\tableaurule}}
\def\gettableau#1 {\ifnum#1=0\let\next=\null\else
  \squares{#1}\let\next=\gettableau\fi\next}

\def\Fund#1#2{\vcenter{\vbox{\drawbox{#1}{#2}}}}
\def\Asym#1#2{\vcenter{\vbox{\drawbox{#1}{#2}
              \kern-#2pt       
              \drawbox{#1}{#2}}}}
\def\fund{\Fund{6.5}{0.4}}
\def\asym{\Asym{6.5}{0.4}}
\def\sym{\tableau{2}}

\input epsf
\newcount\figno
\def\fig#1#2#3{
\par\begingroup\parindent=0pt\leftskip=1cm\rightskip=1cm\parindent=0pt
\baselineskip=11pt
\global\advance\figno by 1
\epsfxsize=#3
\centerline{\epsfbox{#2}}
\vskip 12pt
{\bf Figure \the\figno:} #1\par
\endgroup\par
}
\def\figlabel#1{\xdef#1{\the\figno 
\mbox{ }}}
\def\encadremath#1{\vbox{\hrule\hbox{\vrule\kern8pt\vbox{\kern8pt
\hbox{$\displaystyle #1$}\kern8pt}
\kern8pt\vrule}\hrule}}

\begin{document}

\preprint{
 HUB-EP-97/92\\
 {\tt hep-th/9712143}\\
}

\title{Branes at Orbifolds versus Hanany Witten in Six Dimensions}
\author{Ilka Brunner and Andreas Karch
 \oneaddress{
  \\
  Humboldt-Universit\"at zu Berlin\\
  Institut f\"ur Physik\\
  Invalidenstra{\ss}e 110\\
  D-10115 Berlin, Germany\\
  {~}\\
  \email{brunner@qft1.physik.hu-berlin.de\\
  karch@qft1.physik.hu-berlin.de}
 }
}
\date{December 15, 1997}

\Abstract{
We reconstruct non-trivial 6d theories obtained by Blum and
Intriligator by considering
IIB or $SO(32)$ 5 branes at ALE spaces in the language of Hanany Witten setups.
Using ST duality we make the equivalence of the two approaches
manifest, thereby uncovering several new T-duality relations
between the group theoretic data describing the embedding of the instantonic
5 brane in the ALE and brane positions in the Hanany Witten language.
We construct several new 6d theories, which can be understood
as arising on 5 branes in IIB orientifolds with oppositely charged
orientifold planes recently introduced by Witten. 
}

\maketitle

\section{Introduction}

The past few years have seen tremendous progress in our understanding
of field and string theories. Recent interest has focused
on trying to understand the interconnection
between the various results in field and string theories. One
of the tools used in these studies is the realization of
Super-Yang-Mills theories as the low-energy theory on the 
worldvolume of D-branes. Many strange field theory
phenomena like dualities and non-trivial fixed points
this way find there natural place in string theory. 

Contact between D-brane worldvolumes and stringtheory
has been made by using `branes as probes' \cite{probe,seiberg}. One studies the
worldvolume of a Dp brane in a given geometric background.
Thereby one engineers a p+1 dimensional field theory. 
More
recently Hanany and Witten provided another brane setup \cite{hw} where
interesting dynamics on the worldvolume is engineered by suspending
the Dp branes between two NS branes in the background of various other
branes. Since in this setup one worldvolume
direction is compact, we are
effectively dealing with a p dimensional SYM theory at low
energies.

We'd like to show the relation between these two approaches in 
the case of non-trivial fixed points with $N=1$ in 6 dimensions. The
most prominent example of this kind of fixed point is the small $E_8$
instanton \cite{ganor}. But soon after its discovery it became clear that
there are more of these fixed points. Their existence
was conjectured from gauge theory considerations \cite{seiberg} and 
also constructed on 5 brane probes at orientifolds \cite{seiberg}. Many
were explicitly constructed via F-theory on Calabi-Yau surfaces \cite{vafa}.
In \cite{intblum} even more new 6d fixed points were constructed
by considering 5 brane probes at ADE orbifolds. They arise
as strong coupling limits of certain product gauge groups.

In \cite{d6} a method was explained to obtain several of these
fixed points from a simple Hanany-Witten like setup involving
6 branes and 5 branes. The setup was generalized in
\cite{zaff} by the inclusion of 8 branes. In this paper
we expand the method of \cite{d6} to describe the strong coupling
fixed points of product gauge groups. We show the relation
between this approach and the `5 brane at orbifolds' approach
of \cite{intblum}, thereby uncovering some interesting new T-duality
relations between brane positions and instanton data. In addition
we find some new fixed points which were not realized in a stringy setup
before. They also have an interpretation as `5 branes at orbifolds' once
one allows for orientifolds of IIB theory with oppositely
charged orientifold planes like they were considered recently in
\cite{wittenorbi}.

With the same setups one can also construct `little string theories'
with $N=1$ in the spirit of \cite{little} .

In Section 2 we will review the method of \cite{d6} to construct
$N=1$ 6d fixed points from a Hanany-Witten setup. Section 3 contains
a summary of the results of \cite{intblum}, where the strong coupling
fixed points of certain product gauge groups associated to Dynkin
diagrams are constructed via type IIB and heterotic 5 branes at orbifolds.
In Section 4 we construct the same fixed points via the Hanany-Witten
configuration. We argue that the 2 approaches are in fact closely related.
They can be mapped into each other via an ST duality transformation.
By matching the obtained fixed point theories we uncover some interesting
new T-duality mappings between the group theoretic data describing
the embedding of the instantonic 5 brane in the ALE space
and the position of branes \footnote{This analysis is in a similar
spirit to the one from \cite{singular} where ST duality maps
the data describing monopoles on ALF spaces to brane positions in
the Hanany-Witten language}.

In Section 5 we use this method to construct new fixed point theories and
hence also new little string theories. They can
be obtained in the TS dual language as wordvolume theories of
5 branes in IIB orientifold theories with oppositely charged
orientifold planes introduced recently in \cite{wittenorbi}.

\section{6d Fixed Points from Hanany-Witten Setups}

The original Hanany-Witten \cite{hw}
setup yields an $N=4$ supersymmetric theory in three
dimensions. Its basic ingredients are IIB NS 5 branes, D5 branes and
D3 branes. The worldvolumes of the various branes occupy the
following directions:

\begin{center}
\vspace{.6cm}
\begin{tabular}{|c||c|c|c|c|c|c|c|c|c|c|}
\hline
&$x^0$&$x^1$&$x^2$&$x^3$&$x^4$&$x^5$&$x^6$&$x^7$&$x^8$&$x^9$\\
\hline
NS 5&x&x&x&x&x&x&o&o&o&o\\
\hline
D 5&x&x&x&o&o&o&o&x&x&x\\
\hline
D 3&x&x&x&o&o&o&x&o&o&o\\
\hline
\end{tabular}
\vspace{.6cm}
\end{center}

In the $x^6$ direction the 3 branes are suspended between 5 branes. The
effective low energy theory is the 2+1 dimensional field theory
living on the extended 3 brane directions. Fluctuations of the
larger branes are considered much heavier than those of the 3 branes.
Moving around the higher dimensional branes hence corresponds to
changing parameters of the low energy theory, moving around the
3 branes corresponds to changing moduli of the theory. The
idea behind this approach can be summarized as the statement, that
the low energy dynamic is determined by the lowest dimensional brane in
the setup.
The number $N_c$ of 3 branes determines the gauge group, which
is $U(N_c)$ in the standard setup\footnote{
For 4 and higher dimensions the $U(1)$ part of
the $U(N_c)$ expected on coinciding branes is projected out \cite{witten}}
, or $SO(N_c)$ and $Sp(N_c)$
\footnote{$Sp(N_c)$ denotes for us the group with $N_c$ dimensional
representation. Hence in this case $N_c$ has to be even}
upon including orientifolds. Hence we will refer to them as
color branes. Similarly the D5 act as flavor branes since they
introduce massless hypermultiplets. Another way to
add matter is to add semi-infinite 3 branes.

In \cite{d6} a similar setup was proposed to describe
6d gauge theories with $N=1$ supersymmetry
and their strong coupling fixed points.
Roughly speaking one applies 3 T-dualities to the Hanany-Witten
setup, yielding IIA 6 branes suspended
between NS 5 branes.
The flavor branes in this setup are D8 branes.
These D8 branes only appear once one allows a cosmological constant
\cite{zaff} and hence considers a massive IIA \cite{romans}
configuration.

\subsection{The Basic Brane Configuration}

After applying T-dualities on the 3, 4 and 5 directions we obtain
IIA and the following setup of branes:
two NS 5 branes as usual along the 012345 coordinates. Stretched between
them $N_c$ D6 branes with a worldvolume along the 0123456 coordinates.
In addition there are $n_l$ semi-infinite 6 branes ending
on the left NS brane and $n_r$ semi-infinite 6 branes ending
on the right. We will again call them the $N_f=n_l+n_r$ flavor
branes. The inclusion of D8 branes following the
proposal of \cite{zaff} will be discussed later on.
\vspace{1cm}
\fig{The brane configuration under consideration, giving rise
to a 6 dimensional field theory. Horizontal lines represent
D6 branes, the crosses represent NS 5 branes.}
{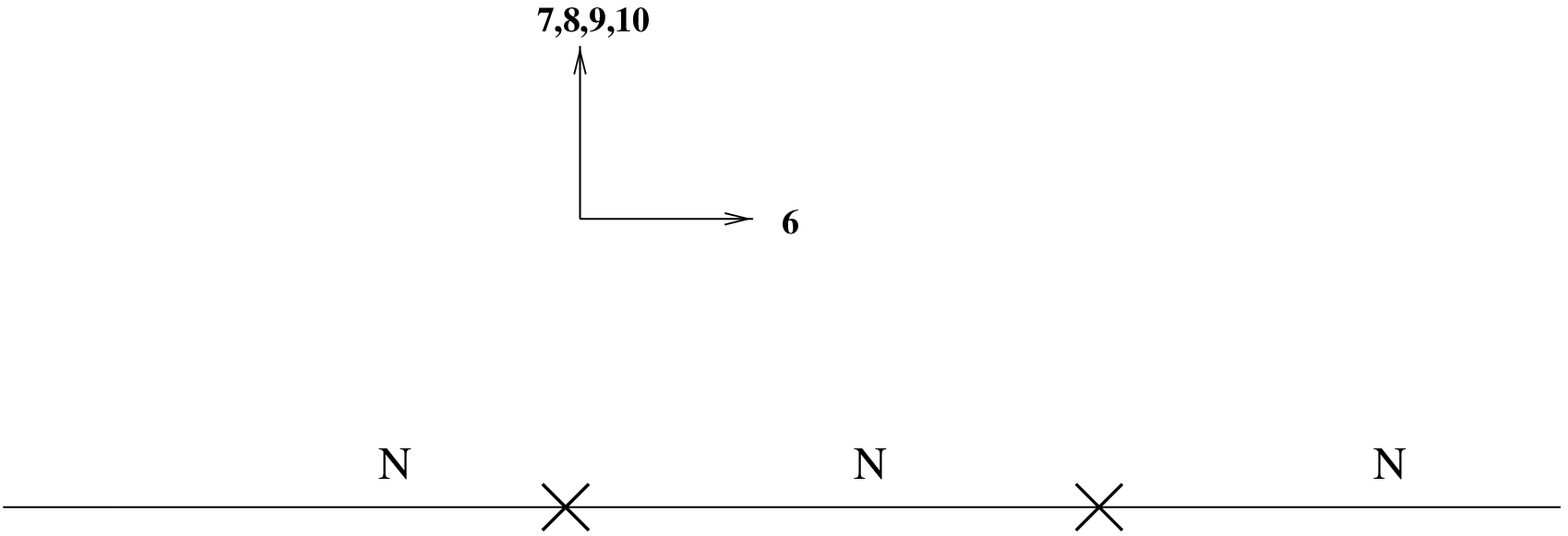}{10truecm}
\figlabel{\basic}
\vspace{1cm}
Figure \basic shows the basic
brane setup for an $SU(N)$ gauge group with $2 N$ flavors.
Note that in order to get 
the low energy field theory corresponding to this brane
setup we should look for the lowest dimensional objects.
These are as usual the finite color branes, but this time we get
an equally important contribution from the full NS branes.
As shown in \cite{d6} every 5 brane contributes one $N=1$ tensor
multiplet. These tensors come from the (2,0) tensor
multiplet on the NS5 brane, which decomposes
to an (1,0) tensor and a (1,0) hyper.
The hyper corresponds to motions of the 5 brane away from the D6 branes
and hence is frozen out from the low energy dynamics \cite{d6}.
One of those tensor multiplets describes the
center of mass motion and hence decouples.

\subsection{Ramond Charge Conservation versus Anomalies}
A D6 brane ending on a NS5 brane is a source of RR 7-form charge.
In lower dimensional Hanany-Witten setups the corresponding
flux is usually absorbed by fields living on the NS5 brane,
leading to a bending of the 5 brane in the worldvolume directions
transverse to the end of the color brane. Since here the
whole 5 brane is the end of the D6 brane we find that
the brane configuration is consistent if and only if
the total RR flux on a given NS brane vanishes. With the
elements discussed so far this is only possible if we have
the same number of D6 branes ending from the left and from the right
on a given NS brane, like it is the case in the $SU(N)$ gauge
theory with $2N$ flavors from Figure \basic. Similar one
can find consistent $SO$ and $Sp$ gauge theories
by introducing orientifolds in the picture.

From the 6d field theory point of view this consistency condition
turns up as an anomaly.
The anomaly arising from vector and hypermultiplets is
\begin{equation} \label{anomaly}
I=\alpha \tr F^4 + c (\tr F^2)^2
\end{equation}
where $\tr$ is the trace in the fundamental representation.
In the case $\alpha = 0, c>0$ the anomaly can be cancelled by
introducing a tensor multiplet.
These anomaly polynomials were analyzed systematically by \cite{sweden}. 
It turns out that the only anomaly free theories involving
only fundamental matter are those seen in the brane picture
\footnote{There are some subtleties for $SU(2)$ and $SU(3)$.
They have no independent 4th order Casimir. There $\alpha$
vanishes automatically and we only get an upper bound on the
number of flavors. However global anomalies \cite{vafa}
in these cases restrict us to $N_f=4,10$ for $SU(2)$ and
$N_f=0,6,12$ for $SU(3)$. The 10 flavor case for $SU(2)$ can be realized
by viewing $SU(2)$ as $Sp(2)$ and include orientifolds.}.
In addition there are anomaly free $SU(N)$ theories with a symmetric
tensor and $N-8$ fundamentals or an
antisymmetric tensor and $N+8$ fundamentals. Their brane realization will be
discussed in Section 4.

The effective coupling of the gauge theory is
$$\frac{1}{g^2_{eff}}=\frac{1}{g_0^2}+ \sqrt{c} \Phi $$
where $\Phi$ denotes the vev of the scalar in the tensor multiplet.
This is a consequence of coupling the tensor to the gauge fields
in order to cancel the anomaly. By redefining the origin of $\Phi$
we can write this as
$$\frac{1}{g^2_{eff}}=\sqrt{c} \Phi. $$
In the brane picture this effective coupling corresponds to the
distance between the two 5 branes. For $\Phi=0$, that is when
the 5 branes coincide, we get a strong coupling fixed point \cite{seiberg}.

It is clear that by repeating the same picture by adding more 5 branes
one creates anomaly free product gauge groups, similar like in 4d 
\cite{brodie}. Every new
5 brane introduces another $SU(N)$ gauge factor, a new tensor multiplet
and a bifundamental hyper multiplet. By tuning the scalars
in all these tensor multiplets to bring all the 5 branes together in one
point each of these product gauge groups exhibits
a non-trivial fixed point.

\subsection{Inclusion of 8 branes}

As discusses above, the general HW setup contains in addition to
the color branes also flavor branes that yield an alternative
way of realizing fundamental hypermultiplets (so far we included
hypermultiplets by semi-infinite color branes). In our case
these flavor branes are D8 branes, as introduced by
\cite{zaff}. This is problematic, since
D8 branes are not a solution of standard IIA theory but require
massive IIA \cite{romans}, that is the inclusion of a cosmological
constant $m$. The D8 branes that act as domain walls between regions
of space with different values of $m$. $m$ is quantized and in appropriate
normalization \cite{zaff} can be chosen to be an integer.

The presence of $m$ makes itself known to the brane configuration via
a coupling
\begin{equation}
\label{m}
-m\int dx^{10} B\wedge
*F^{(8)},
\end{equation}
where $B$ is the 2 form NS gauge field under which the NS5 brane is charged
and F is the 8 form field strength for the 7 form gauge field under
which the D6 is charged. 
From the Bianchi identity for the 2 form field strength dual
to $F^{(8)}$ in the presence of a 
D6 brane ending on a 5 brane, $dF^{(2)}=d*F^{(8)}= \theta
(x^7)\delta^{(456)} - mH$ (where $H=dB$), one gets $mdH=\delta^{(4567)}$.
The $\delta$ function comes from the charge of the D6 brane-end.
From here we get back the old result, that for $m=0$ 
D6 branes always have to end on a given NS5 brane in pairs
of opposite charge (that is from opposite sides). However for arbitrary 
$m$ this relation shows that RR charge conservation requires that
we have a difference of $m$ between the number of branes ending from
the left and from the right 
\footnote{Throughout the paper we will use the following conventions
to fix the signs: by passing through an D8 from left to right
$m$ increases by 1 unit. In a background of a given $m$ the number
of D6 branes ending on a given NS5 from the left is by $m$ bigger than
the number of D6 branes ending from the right.}.

\vspace{1cm}
\fig{Basic Hanany Zaffaroni Setup}
{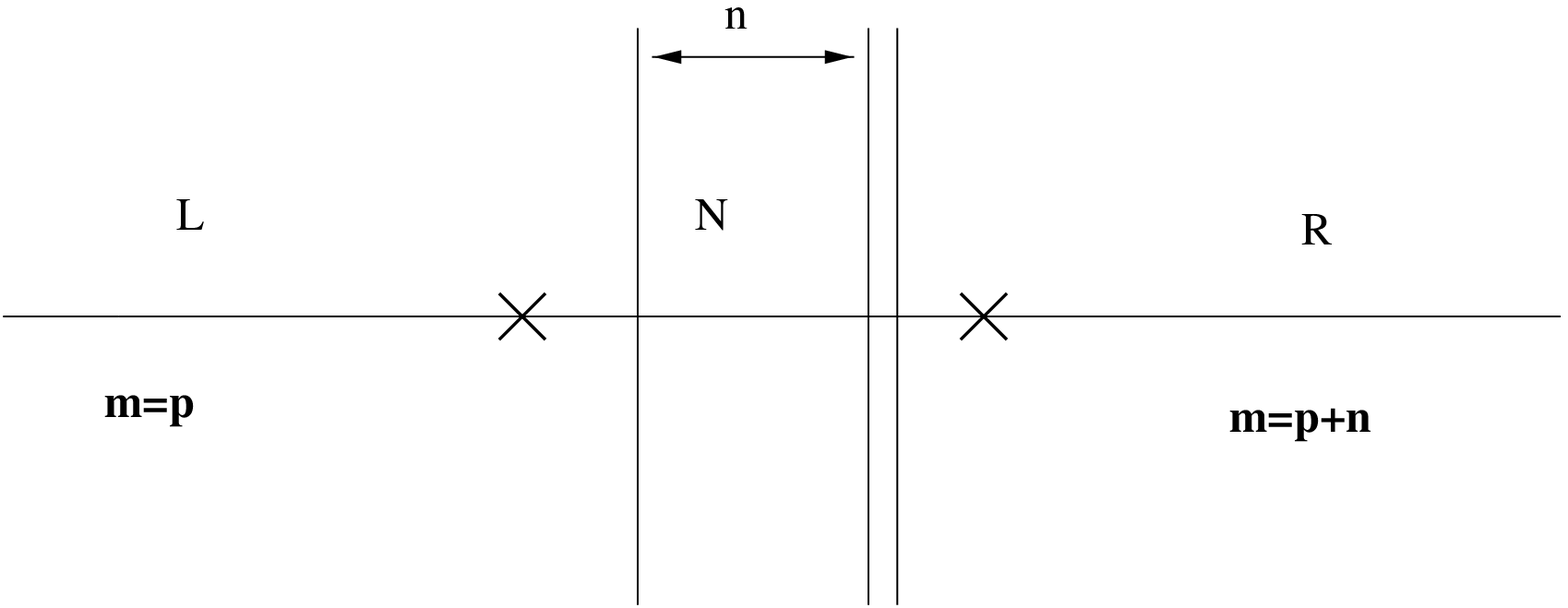}{10truecm}
\figlabel{\han}
\vspace{1cm}

Figure \han shows the basic brane configuration involving D8 branes.
The $n$ D8 branes in the middle give rise to 8 flavors for the
$SU(N)$ gauge group. In addition they raise the cosmological constant
$m$ from $p$ to $p+n$. In the background of this values of $m$
the modified RR charge conservation tells us
\begin{eqnarray*}
N&=& L- p \\
R&=& N -(p+n) = L -2p -n
\end{eqnarray*}
With the $R+L$ flavors from the semi-infinite D6 branes the total number
of flavors is $L+R+n=2N$ in agreement with the gauge anomaly considerations
on the 6 brane \cite{sweden}
for every possible value of $p$.

\subsection{Orientifolds}

To obtain $SO$ or $Sp$ groups we have to introduce orientifold planes
as in \cite{zwieb, shapere, EKG2, theisen}. We have in principle
2 possibilities: O8 branes and O6 branes. Let us first consider the
O8. We have to distinguish two possibilities: O8 planes with negative or
positive D8 brane charge (that is -16 or +16) \cite{polchinski}.
The former are the T-dual of the O9 projecting
IIB to type I. On the D8 worldvolume they project the symmetry group to
an (global) $SO$ group, while on the D6 we get a local $Sp$ group.
The positively charged O8 projects on global $Sp$ and local $SO$.
If we want to have vanishing total D8 charge, we should restrict
ourselves
to either 2 negatively charged O8s with 32 D8 branes or one O8 of each type.
Latter configuration was recently studied by Witten \cite{wittenorbi}.

Since these O8 planes carry D8 brane charge, they also effect the
cosmological constant like -16 (+16) D8 branes would.

Building product gauge groups from these we have the four
possibilities.

\vspace{1cm}
\fig{Various possibilities to introduce O8 planes}
{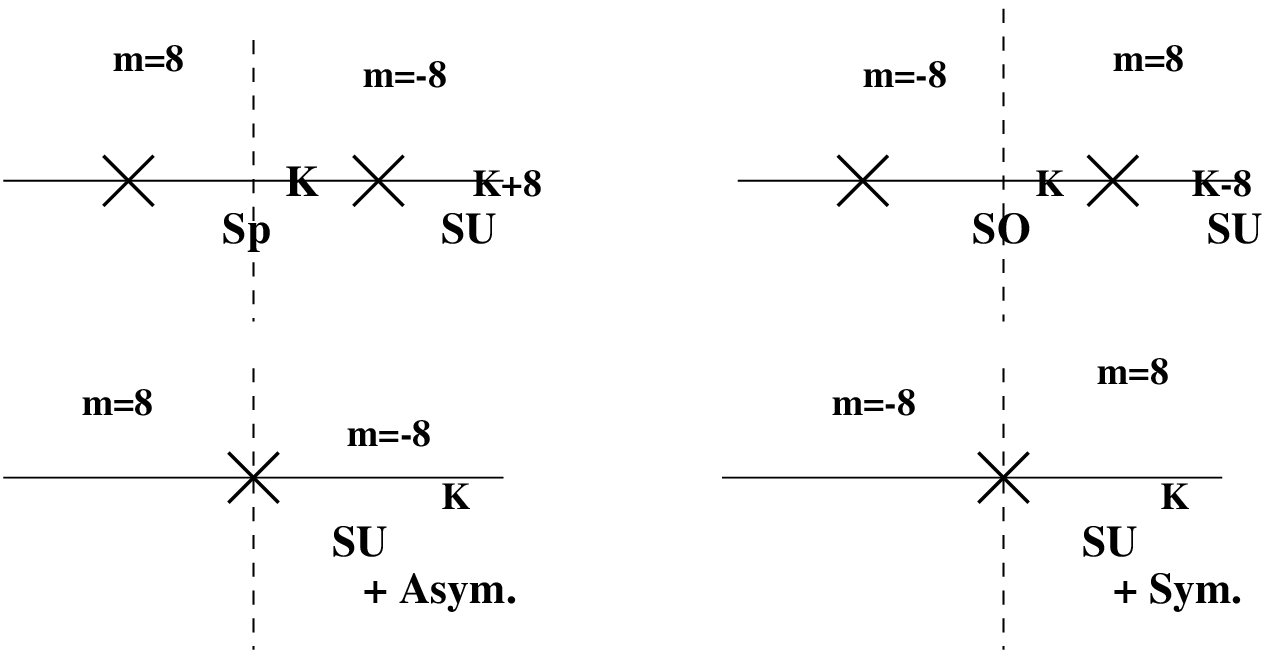}{10truecm}
\figlabel{\orient}
\vspace{1cm}

The gauge groups corresponding to brane setups in Figure \orient have been
analyzed in the equivalent setup for 4d in \cite{curves}.
Their result is indicated in Figure \orient. If the O8 is in between
two 5 branes the `center' 
gauge group is projected to $Sp(K)$ or $SO(K)$ respectively.
All other gauge groups stay $SU$, however the $SU$ groups to the right
are identified with those to the left and one effectively projects out
half of the $SU$ groups. In addition to this, in 6d we got the special
situation that the O8 also changes the cosmological constant. For symmetry
reasons we have to choose $m=\pm 8$ on the two sides of the orientifold.
Therefore we obtain in total
$$ Sp(k) \times \prod_i SU(k+8 i) $$
or
$$ SO(k) \times \prod_i SU(k - 8 i) $$
respectively. From strings stretching in between neighbouring
gauge groups we get bifundamentals $(\fund,\fund)$.

The other possible situation considered in \cite{curves}
is that one of the 5 branes is stuck to the O8. In this case
all the groups stay $SU$, again the left and the right ones identified
leading to effectively half the number of gauge groups. In addition
the middle gauge group has a matter multiplet in the antisymmetric
or symmetric representation for positive/negative charge O8 planes.
In 6d we again have the effect of the cosmological constant
and as a result get
$$ \prod_i SU(k \pm 8i)$$
with antisymmetric/symmetric tensor matter in the first gauge group factor
in addition to the bifundamentals. Note that these gauge theories
are all anomaly free \cite{sweden}
upon coupling to the tensor multiplets associated
to the independent motions of the 5 branes.

The other realization of $SO$ and $Sp$ groups is to introduce
an O6 along the D6 branes. Here the discussion in 6d is
analog to the original one in 4d with O4 orientifolds
\cite{shapere,EKG2} and
was performed in \cite{d6}. Upon building product
gauge groups we get an alternating series of $SO$ and $Sp$
groups with bifundamentals like in 4d \cite{radu}.
As in 4d this is due to the fact that the orientifold changes sign upon
passing through the NS5 brane \cite{shapere}. It therefore also contributes
to the RR charge conservation condition. We obtain a
$$ \prod_i SO(N+8) \times Sp(N)$$
chain with bifundamentals.

\section{6d Fixed Points From 5 Branes at Orbifolds}
In this Section we will review the analysis of \cite{intblum}.
\subsection{$U(N)$ ALE instantons and IIB NS5 Branes at Orbifolds}
6d fixed points and `little string theories' 
with (1,1) or (2,0) supersymmetry arise as worldvolume
theories on an NS5 brane of IIB and in the 5+1 dimensional
space transverse to an ADE singularity respectively,
that is one looks at IIB on $\IC ^2/\Gamma _G$ where $\Gamma_G$
is a discrete $SU(2)$ subgroup with ADE classification.

The main idea of \cite{int,intblum,intstring} was to combine
these and study NS5 branes at ADE singularities to obtain new
$N=1$ (that is (1,0)) supersymmetric theories. 
One can determine the gauge group corresponding
to this setup in the spirit of the analysis of \cite{moore}
of instantons on ALE spaces
\footnote{$K$ p branes inside $N$ p+4 branes can be
interpreted as $K$ $U(N)$ instantons \cite{doug}. The gauge
group on the p branes is given by the
Hyper Kahler quotient construction of instanton
moduli spaces by Kronheimer \cite{kronh}. Since IIB does
not allow for 9-branes, we have $N=0$ and hence the
interpretation in terms of instantons breaks down.
Nevertheless the analysis of the gauge theory on
the 5 brane goes through even in the $N=0$ case.}
to be
\begin{equation}
\prod_{\mu=0}^r U( K n_{\mu})
\label{iibgg}
\end{equation}
where $\mu =0\dots
r=$rank$(G)$ labels the nodes
of the extended Dynkin diagram of the $ADE$ group $G$
corresponding to the irreducible 
representations $R_\mu$ of $\Gamma _G$,
with $|R_\mu|=n_\mu$, the Dynkin indices.
The matter multiplets transform as
$\half \oplus _{\mu
\nu}a_{\mu\nu}(\fund _\mu , \overline {\fund _\nu})$,
where $a_{\mu \nu}$ contains the information about the links in
the extended Dynkin diagram, that is
$a_{\mu
\nu }$ is one if nodes $\mu$ and $\nu$ are linked in the extended
Dynkin diagram and zero otherwise.
In addition,
there are $r$ ${\cal N}=(1,0)$ hypermultiplets and $r$ ${\cal
N}=(1,0)$ tensor multiplets
(which combine into $r$ ${\cal N}=(2,0)$
matter multiplets), coming from reducing the 10d two-form and
four-form potentials 
down to 6d on the $r$ cycles which generate $H_2$
of $\IC ^2/\Gamma _G$.
This theory is anomaly free. The deadly $F^4$ term vanishes
and the remaining gauge anomaly is cancelled by the coupling
to the tensors.

The diagonal $U(1)$ factor in $\prod _{\mu =0}^rU(Kn_\mu)$ has no
charged matter and decouples.  The other $r$ $U(1)$ factors do have
charged matter and are thus anomalous.
This anomaly is cancelled \cite{intblum} with help of the
$r$ hypermultiplets from above, which
correspond
to the blowing up modes of the $\IC
^2 /\Gamma _G$ singularity.
They pair up with the
$U(1)$ gauge fields to give them a mass and their 
expectation values
effectively become \FI\ parameters in the gauge group.
The cancellation of the $U(1)$ anomalies using the $r$
hyper-multiplets and the cancellation of the $F^2$
anomalies using the $r$ tensor multiplets can be
regarded as being related by a remnant of the larger supersymmetry of
the full type IIB theory.

Following \cite{seiberg},  there can be a non-trivial 6d RG
fixed point at the origin of the Coulomb branch 
provided all $g_{\mu,
eff}^{-2} (\Phi )$ ($\Phi$ denotes the scalar in the tensor multiplet) are 
non-negative along some entire ``Coulomb
wedge''. However \cite{intblum} showed that there
is always one subgroup for which $g_{\mu,
eff}^{-2}$ must become negative for large $\Phi$.
As already pointed out in \cite{int}, one
can always take the
$U(K)$ corresponding to 
the extended Dynkin node $\mu =0$ to be the IR
free theory, which means 
that this gauge group is un-gauged in the IR
limit and hence becomes a global symmetry.
It is then possible to choose a Coulomb wedge so that the
remaining gauge groups in (\ref{iibgg}) all have $g_{\mu, eff} ^{-2}(\Phi
)\geq 0$ along the entire wedge even in the $g_{\mu
,cl}^{-2}\rightarrow 0$ limit.

With this the result of \cite{intblum} is that
for any $\IC ^2/\Gamma _G$, and for every $K$, these theories
give 6d non-trivial RG fixed points with $r$ tensor multiplets and
gauge group $\prod_{\mu =1}^rSU(Kn_\mu)$,
 where the $U(1)$s in (\ref{iibgg})
have been eliminated, as discussed above, 
by the anomaly cancellation
mechanism, and the $\mu =0$ node gives a global rather than local
$SU(K)$ symmetry.

If one is considering the `little string theories' positivity
of $g_{\mu, eff} ^{-2}$ is not required. Negative
 $g_{\mu, eff} ^{-2}$ just means that the theory hits
a Landau pole and hence needs new UV degrees of freedom to
be well defined. These additional degrees of freedom are
provided by the `little string'. Hence all $SU(K)$ factors
stay interacting. The $U(1)$s and the additional
hypermultiplets still drop out via the anomaly cancellation
mechanism. 

\subsection{$SO(32)$ ALE instantons}

Another way to obtain new $N=1$ theories is to study
heterotic $SO(32)$ or type I 5 branes at ADE singularities.
This was also done by \cite{intblum}. We can view these
theories as IIB orientifolds with one O9 and 32 D9 branes.
As mentioned above, $K$ 5 branes inside $N$ 9 branes
can be interpreted as $K$ $U(N)$ instantons, and
with the the O9 orientifold $K$ 5 branes are $K$
$SO(N)$ instantons. The corresponding
gauge group is again obtained via the Hyper Kahler
quotient construction \cite{kronh,moore}. While
this construction works for arbitrary $N$ the gauge
theory on the 5 brane is 
anomalous unless $N$ takes the physical
values $N=0$ for $U(N)$ (that is the pure type IIB we
discussed in the previous subsection) and $N=32$ for $SO(N)$.
In addition one has to trade 29 hypermultiplets of the
Hyper Kahler construction for each of the tensor multiplets
that appear in the gauge theory: while the Higgs branch
of the theory is supposedly equivalent to the moduli space of
instantons given by the Hyper Kahler quotient, the
gauge theory on the worldvolume is actually the Coulomb
branch of the small instanton. Making this transition
exactly trades the 29 hypers into 1 tensor.

Putting the 5 branes at the ADE singularity means that we are
dealing with $SO(32)$ instantons on an ALE space. Such instantons
are not only characterized by $K$, but there can
also be non-trivial Wilson lines at infinity \cite{intblum}.
An instanton gauge connection is
asymptotically flat and thus usually trivial since the asymptotic
space $X_\infty$ surrounding the instanton usually has trivial $\pi
_1$.  However, on the ALE space, $\pi _1(X_{\infty})=\Gamma _G$ and
thus there can be non-trivial Wilson lines at infinity, leading to
non-trivial group elements 
$\rho_{\infty} \in  \G$, representing $\Gamma_G$
in the gauge group. 
Thus, in addition to $K$, the instanton is
topologically characterized by integers $w_\mu$ in $\rho _\infty
=\oplus _\mu w_\mu R_\mu$, giving the representation of $\Gamma _G$ in
$\G$ in terms of the irreps $R_\mu$.
In order to have $\rho_{\infty} \in SO(32)$ we must have
$\sum_{\mu} n_{\mu} w_{\mu} =32$ and $w_{\mu} = w_{k+1-\mu}$.

In addition there is another discrete choice one can make
for $\rho _\infty$, due
to the fact that the gauge group of the heterotic string
is in fact $Spin(32)/\IZ_2$ and not $SO(32)$. The 
$\IZ _2$ is generated by the element $w$ in the center
of $Spin(32)$ which acts as $-1$ on the vector, $-1$ on the spinor of
negative chirality, and $+1$ on the spinor of positive
chirality. Because only representations with $w=1$ are in the
$Spin(32)/\IZ _2$ string theory, the identity element $e\in \Gamma _G$
can be mapped to either the element $1$ or $w$ in $Spin(32)$.
Using the terminology of \cite{int} we will refer to this
as the case with or without vector structure
\footnote{One is really only talking about the behaviour
of a flat connection at infinity. There can be non-trivial
obstructions to vector structure even when the
flat connection at infinity is trivial \cite{aspinwall}}.
The integers $w_{\mu}$ and the distinction of with or without
vector structure will have an interpretation in terms
of brane positions in the dual HW setup.

\subsubsection{The case with vector structure}
Consider the worldvolume theory of $K$ type
I 5 branes interpreted as instantons on an ALE space.
The non-trivial Wilson lines are described by the integers
$w_{\mu}$ with $\sum_{\mu} n_{\mu} w_{\mu} =32$.
The analysis for an arbitrary ADE-type singularity
was performed in \cite{intblum}. We will just state their
results for the A$_k$ (that is $\Gamma=\IZ_{k+1}$) singularity, which already
appeared earlier in \cite{int}. For a detailed discussion
of this result we refer the reader to \cite{intblum}.

For $k+1$ {\bf even} the gauge group is
\begin{equation}
\label{evenwith}
Sp(V_{0}) \times \prod_{\mu=1}^{(k-1)/2} SU(V_{\mu}) \times \; Sp(V_{(k+1)/2})
\end{equation}
with $\half w_0 \fund_0$, $\oplus_{\mu=1}^{(k-1)/2} w_{\mu} \fund_{\mu}$,
$ \half w_{(k+1)/2} \fund_{(k+1)/2}$ and $ \oplus_{\mu=1}^{(k+1)/2}
(\fund_{\mu-1}, \fund_{\mu})$ matter multiplets (subscripts label
the gauge group). In addition there are $(k+1)/2$ tensor multiplets.
The $V_{\mu}$ are given by
$$V_0=2K, \; \; \; V_{i \neq 0} = \sum_{j=1}^k C^{-1}_{ij} (w_j + D_j)$$
Where $C^{-1}_{ij}$ is the inverse $SU(k+1)$ Cartan matrix,
given by $C^{-1}_{i<j}=i (k+1-j)/(k+1)$. $w_{\mu}$ are
defined above to contain the information about the Wilson lines
at infinity. While in the Hyper Kahler quotient construction
$D_{\mu}$ is given by $D_{\mu}=- \delta_{\mu,0}$, anomaly freedom
of the gauge theory demands $D_{\mu} =- 16 \delta_{\mu,0} -16 \delta_{\mu ,
(k+1)/2}$. This corresponds precisely to the trading of the
hypermultiplets on the Higgs branch to the tensors on the Coulomb branch.
With this information one can write the $V_{\mu}$ as
$$
 V_{\mu} = 2K + \sum_{\nu=0}^{(k+1)/2} min(\mu,\nu) \cdot
 W_{\nu} - 8 \mu $$
where $W_0 = \half w_0$, $W_{(k+1)/2}= \half w_{(k+1)/2}$ and all other
$W_{i}=w_{i}$.

\vspace{1cm}
For $k+1$ {\bf odd} we similarly get
\begin{equation}
\label{odd}
Sp(V_{0}) \times \prod_{\mu=1}^{k/2} SU(V_{\mu})
\end{equation}
with $\half w_0 \fund_0$, $\oplus_{\mu=1}^{k/2} w_{\mu} \fund_{\mu}$, $
 \oplus_{\mu=1}^{k/2}
(\fund_{\mu-1}, \fund_{\mu})$ and
$\asym_{k/2}$ matter multiplets 
and $k/2$ tensor multiplets.
Similar as above the $V_{\mu}$ are given as
\footnote{In this case $D_{\mu} = -16 \delta_{\mu,0} -8 \delta_{\mu,
k/2} - 8 \delta_{\mu, (k+2)/2}$.}
$$ V_{\mu} = 2K + \sum_{\nu=0}^{k/2} min(\mu,\nu) \cdot
 W_{\nu} - 8 \mu $$
where this time only $W_0 = \half w_0$ and all other
$W_{i}=w_{i}$. This identification will become more
transparent in the HW picture.

\subsubsection{The case without vector structure}
For the case without vector structure only the A-type
singularity was analyzed in \cite{int}. This possibility
only exists for $k+1$ even. The result is
\begin{equation}
\label{evenout}
\prod_{\mu=0}^{(k-1)/2} SU(V_{\mu})
\end{equation}
with $ \oplus_{\mu=1}^{(k-1)/2} w_{\mu} \fund_{\mu},
 \oplus_{\mu=0}^{(k-1)/2}
(\fund_{\mu-1}, \fund_{\mu}), \asym_{0}$ and
$\asym_{(k-1)/2}$ matter multiplets
and $(k-1)/2$ tensor multiplets.
The $V_{\mu}$ are in this case
$$ V_{\mu} = 2K + \sum_{\nu=0}^{(k-1)/2} min(\mu,\nu) \cdot
 w_{\nu} - 8 \mu. $$
Where in this case on had to use 
$D_{\mu} = -8 \delta_{\mu,0} -8 \delta_{\mu,(k-1)/2}
-8 \delta_{\mu,(k+1)/2}-8 \delta_{\mu,k}$.

\subsection{Theories with Twisted 5 Branes}
Another class of fixed points has been found in \cite{intblum}
by considering what they called twisted 5 branes in type I.
Again we will just quote their result for the A$_k$ case. The
interpretation of the twisted 5 brane will become clear by
considering the dual HW setup.
They only exist for even $k+1$.
The theory obtained this way for A$_k$ singularity and $K$ NS branes has
gauge group
$$
\prod_{s=0}^{(k-1)/2} Sp(V_{2s}) \times SO(V_{2s+1})
$$
and  $\oplus_{\mu=1}^{k+1}
(\fund_{\mu-1}, \fund_{\mu})$ matter multiplets (where the $(k+1)$th
gauge group is identified with the 0th).
Similar to the other cases 
$$V_0=2K, \; \; \; V_{i \neq 0} = \sum_{j=1}^k C^{-1}_{ij} D_j$$
and $D_{\mu}=(-16,16,-16,\ddots,-16,16)$. It's easy to see that
this yields
\begin{equation}
\label{twist}
\left \{ Sp(2K) \times SO(2K+8) \right \}^{(k+1)/2} 
\end{equation}
The gauge anomaly is cancelled by $k$ tensors.

\section{Relation between the Two Approaches}

The fixed points described in the last section arise as the worldvolume
theories of NS or heterotic 5 branes. In the spirit of \cite{seibergsethi}
one would not expect these probes to be good probes of the background geometry.
In the limit that we get a decoupled theory on the brane any
information about compactness of the background is lost.
 Clearly
the singularity type matters since one obviously gets different theories
by considering different singularities. This means that we should see the same
picture by considering 5 branes with a transverse ALF space instead of
the ALE, that is a space that asymptotically looks like
($\IR^3 \times S^1)/ \Gamma$ instead of the ALE which looks like
$\IR^4 / \Gamma$ 
But this configuration can be looked at in a ST dual picture:
S-duality takes the IIB NS5 brane into an D5 brane or similar the heterotic
5 brane into an type I 5 brane, which
can also be considered a IIB D5 at an orientifold 9 plane with 32 embedded
D9. T-duality on the ALF isometry turns
D5 branes into D6 branes and an A$_k$ singularity into $k+1$
NS 5 branes. Note that this exactly maps the description
of Section 3 to that of Section 2. 
This mapping is valid whether one goes to the limit
where the theory on the brane decouples as a `little string'
theory \cite{little,intstring} or if we also decouple
these stringy excitations and study the strong coupling
fixed point. However the dual picture of the `little string'
theories necessary involves a limit where the dual radius
is taken to zero.
Therefore we will mostly limit ourselves
in the rest of the discussions to the realization of
strong coupling fixed points
of the renormalization group.
We will explore this correspondence
in more detail in the various cases thereby uncovering several
interesting T-duality relations.

Note that almost the same duality has been used in \cite{matrix}
to map the QM system describing D0 branes on an ALE space to
a Hanany Witten setup with strings stretching between 5 branes.

\subsection{IIB NS5 Branes at an A$_k$ Singularity}

As shown by \cite{int,intblum} and reviewed above, the gauge theory
on $K$ NS5 branes on a transverse A$_k$ singularity yield an
anomaly free
$SU(K)^{k+1}$ gauge group with $k$ tensor multiplets and bifundamentals
transforming under the $i$th and $(i+1)$th gauge group, where the $(k+2)$th
gauge factor is again the first. Or in the language of
\cite{intblum} we have to use the values
of $n_{\mu}$ and $a_{\mu \nu}$ for the A$_k$ extended
Dynkin diagram, that is $n_{\mu}=1$ for all $\mu$
and only $a_{\mu \nu}= \delta_{\mu,\mu+1}$. With this data we obtain 
\begin{equation}
\prod _{\mu =0}^r U(Kn_\mu )
\label{A}
\end{equation}
gauge group with matter multiplets in representations
$\half \oplus _{\mu
\nu}a_{\mu\nu}(\fund _\mu , \overline {\fund} _\nu)$.
(\ref{A}) reduces to (\ref{iibgg}) once on takes into account that
the additional hypermultiplets and $U(1)$s
have been eaten up by the anomaly cancellation mechanism
of \cite{intblum}.
In order to go to the non-trivial fixed points we will have to tune all
coupling constants to infinity. As mentioned above this is not possible.
One gauge factor always becomes IR free, so that at the fixed point
we only have an $SU(K)^{k}$ gauge group. The last $SU(K)$ factor
becomes a global symmetry.

The same theory is described in the following HW setup with $k$ NS branes.
\vspace{1cm}
\fig{Brane configuration yielding a $SU(K)^{k}$ gauge group with
bifundamentals}
{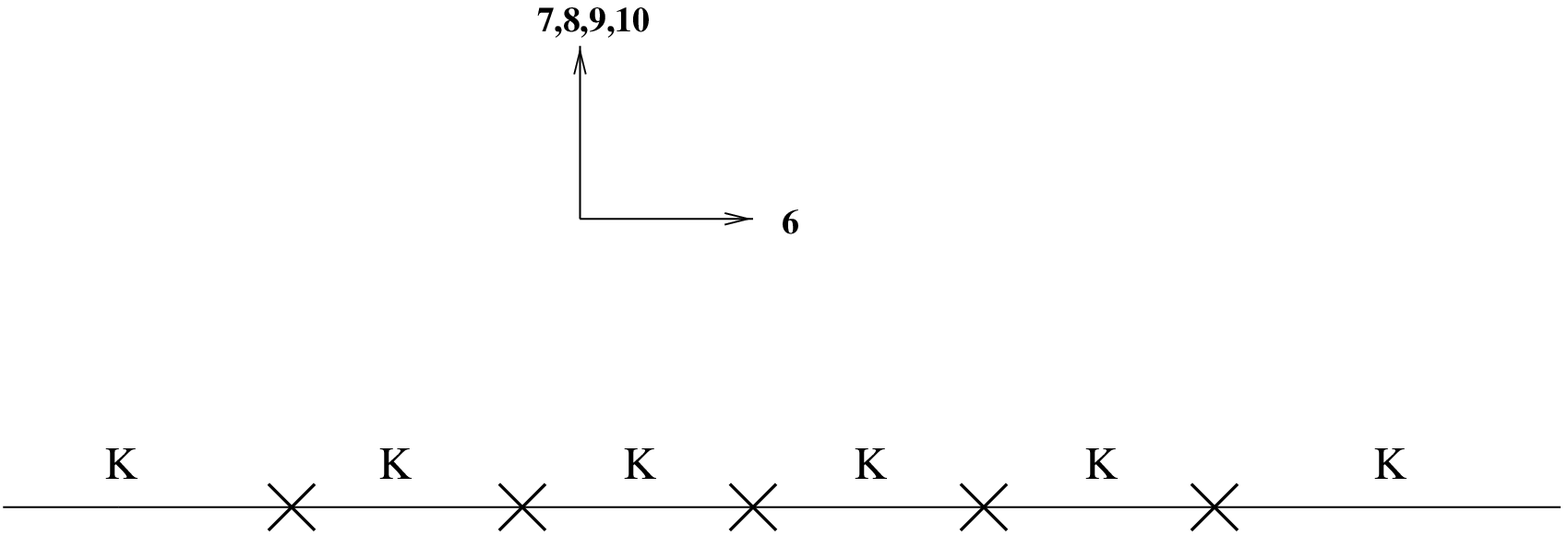}{10truecm}
\figlabel{\su}
\vspace{1cm}
RR charge conservation at every NS5 brane guarantees anomaly freedom
of the gauge theory. The matter content and gauge group can be simply read
off from the brane construction. It agrees with the one from above.

Turning on \FI\ terms in the HW picture corresponds to moving
the 5 branes away in the 7,8,9 (and 10, once we lift to M-theory)
direction. This leaves us with one less 5 brane, corresponding,
as expected, to blowing up the A$_k$ singularity to an A$_{k-1}$.
As reviewed in Section 2 by letting the 6 branes end on the 5 branes
in the HW setup we effectively locked the hypermultiplet arising
from the decomposition of the (2,0) tensor on the NS5 brane under (1,0)
supersymmetry and in addition froze out the $U(1)$ part of the
gauge theory on the D6 brane. This is the dual realization of the
anomaly cancellation mechanism of \cite{intblum}: a hypermultiplet
and a $U(1)$ vector get massive and turn into an \FI\ term for
the gauge theory, corresponding to a blowing-up mode of the singular
space.

To make the duality connection we compactify the 6 direction. Upon TS
duality the $K$ 6 branes turn into $K$ NS branes and the $k+1$ 
NS branes yield the
A$_k$ singularity. We hence made the equivalence of the
two approaches manifest. For various
limits of the string coupling and the radius of the compact dimension
one or the other description is more appropriate.
\vspace{1cm}
\fig{Brane configuration of \protect \su with $x_6$ compact}
{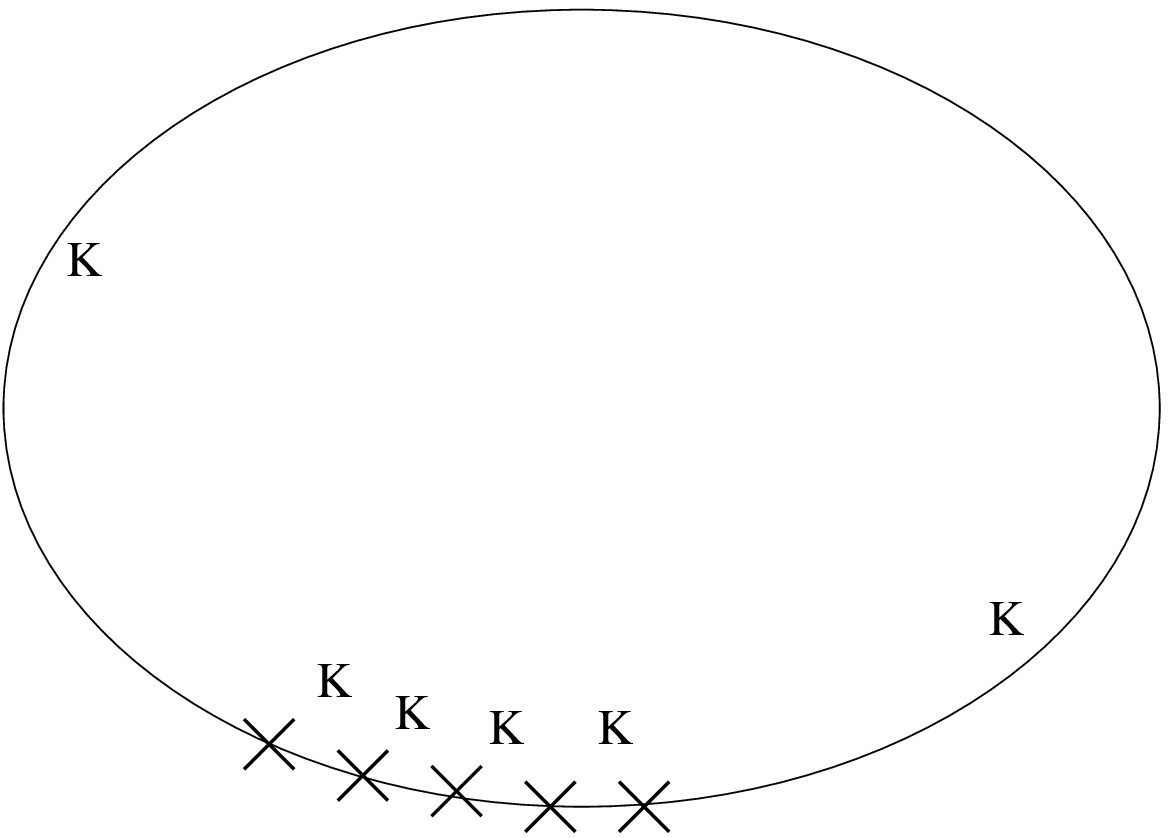}{7truecm}
\figlabel{\su_cp}
\vspace{1cm}
Once we compactified on the circle we see the full $SU(K)^{k+1}$
gauge symmetry. Tuning the moduli in such a way that we obtain
the fixed point theory amounts to moving all the 5 branes on top
of each other. Since the distance between the 5 branes is the
effective coupling constant of the corresponding gauge theory
we see that there is always one gauge group who's color branes
stretch around the entire circle, no matter where we choose to
bring our 5 branes together. This
is a nice realization of the fact that there's always
one gauge factor with
a finite gauge coupling that becomes free in the IR and hence
becomes a global symmetry. 

A special case of this setup on the circle is if we have only one
5 brane on the circle. In this case the bifundamental becomes just
an adjoint. There are no tensor multiplets left. In the field theory
in this case the anomaly
polynomial is identical zero. In fact the adjoint hypermultiplet combines
with the vector multiplet to yield an (1,1) supersymmetric gauge
theory. This is of course no big surprise, since on the
dual side we are in this case just dealing with the theory of IIB NS5 branes
on a non-singular space.

\subsection{$SO(32)$ 5 Branes at an A$_k$ Singularity}

From the discussion of section 3 we recall that
according to the analysis of \cite{intblum} $SO(32)$
5 branes at an A$_k$ singularity can be characterized by
the following data: $K$, the number of 5 branes, the singularity
A$_k$, the Wilson lines at infinity described by integers $w_{\mu}$
satisfying $\sum_{\mu} w_{\mu}= 32$
($\mu$ runs from 0 to $k$ labelling the nodes of the
extended Dynkin diagram)
and the the discrete choice whether the identity in $\Gamma_{k+1}$
is mapped to 1 or to $w$ in $Spin(32)/Z_2$, that
is to say in the language of \cite{int} whether we are considering the
case `with or without' vector structure.

In what follows we will reproduce the same gauge theories in
the language of the Hanany-Witten setup thereby producing
a dictionary of how the data describing the theory gets mapped
under T-duality. Information about the geometry
of the gauge bundle like the integers $w_{\mu}$ and the
distinction of `with or without' vector structure have
a nice interpretation just in terms of brane positions.

\vspace{1cm}
\fig{Brane configuration yielding a product of two $Sp$ and several
$SU$ gauge groups
bifundamentals }
{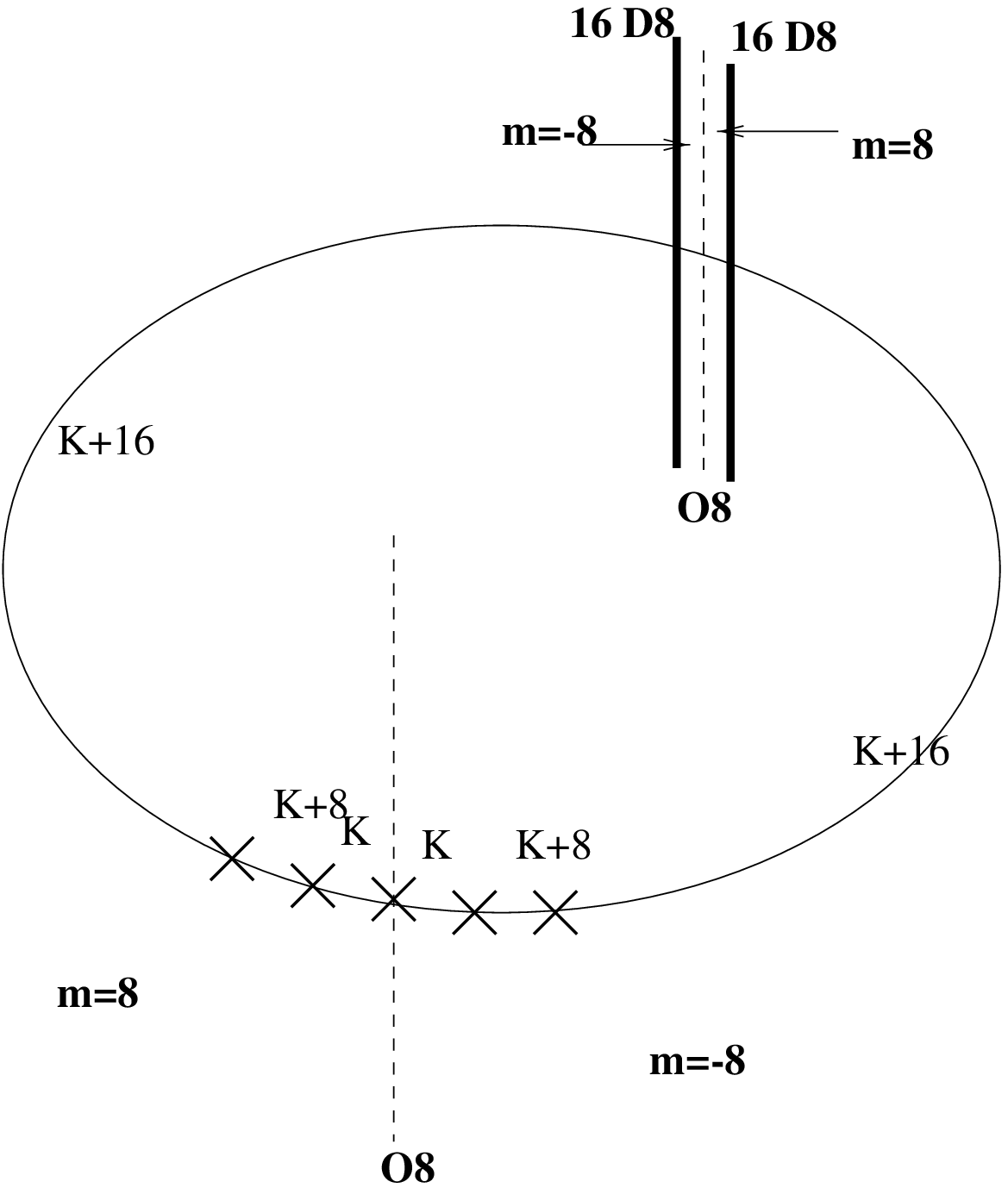}{7truecm}
\figlabel{\sp_cp}
\vspace{1cm}
 Figure \sp_cp shows the basic HW setup describing this kind of
theory. In addition to the $k+1$ NS 5 branes corresponding
to the A$_k$ singularity and the D6 branes stretching between
them, we introduced
two O8 orientifold planes. They are the T-dual picture
of the O9 which one introduces to go from IIB to type I.
In addition we have 32 D8 branes. Their position on the
circle is described by the integers $w_{\mu}$ that contained
the data about the instanton embedding in the picture of 
\cite{intblum}:
$w_{\mu}$ just counts the number of D8 branes between
5 brane number $\mu$ and number $\mu+1$. Similar the $D_{\mu}$
from above which was chosen to have a particular value to cancel
the anomaly corresponds to D8 brane charge from the
orientifolds. The anomaly free
values of $D_{\mu}$ just tell us that the orientifolds
are two lumps of D8 brane charge -16 each on symmetric positions
between the 5 branes.
The condition
$w_{\mu}=w_{k+1-\mu}$ is simply the statement that
the construction is symmetric with respect to the orientifolds.
If $k+1$ is odd, one of the 5 branes is
stuck at one of the orientifold planes. If $k+1$ is
even, we can either have all 5 branes free to move around
on the circle or we have one 5 brane stuck at each of the two
orientifolds. This will correspond to the distinction of
`with or without' vector structure. We will
explain this correspondence in more detail by comparing
the gauge groups that arise.

\vspace{1cm}
\fig{Notation in the example of A$_6$, that is 7 5 branes on the circle}
{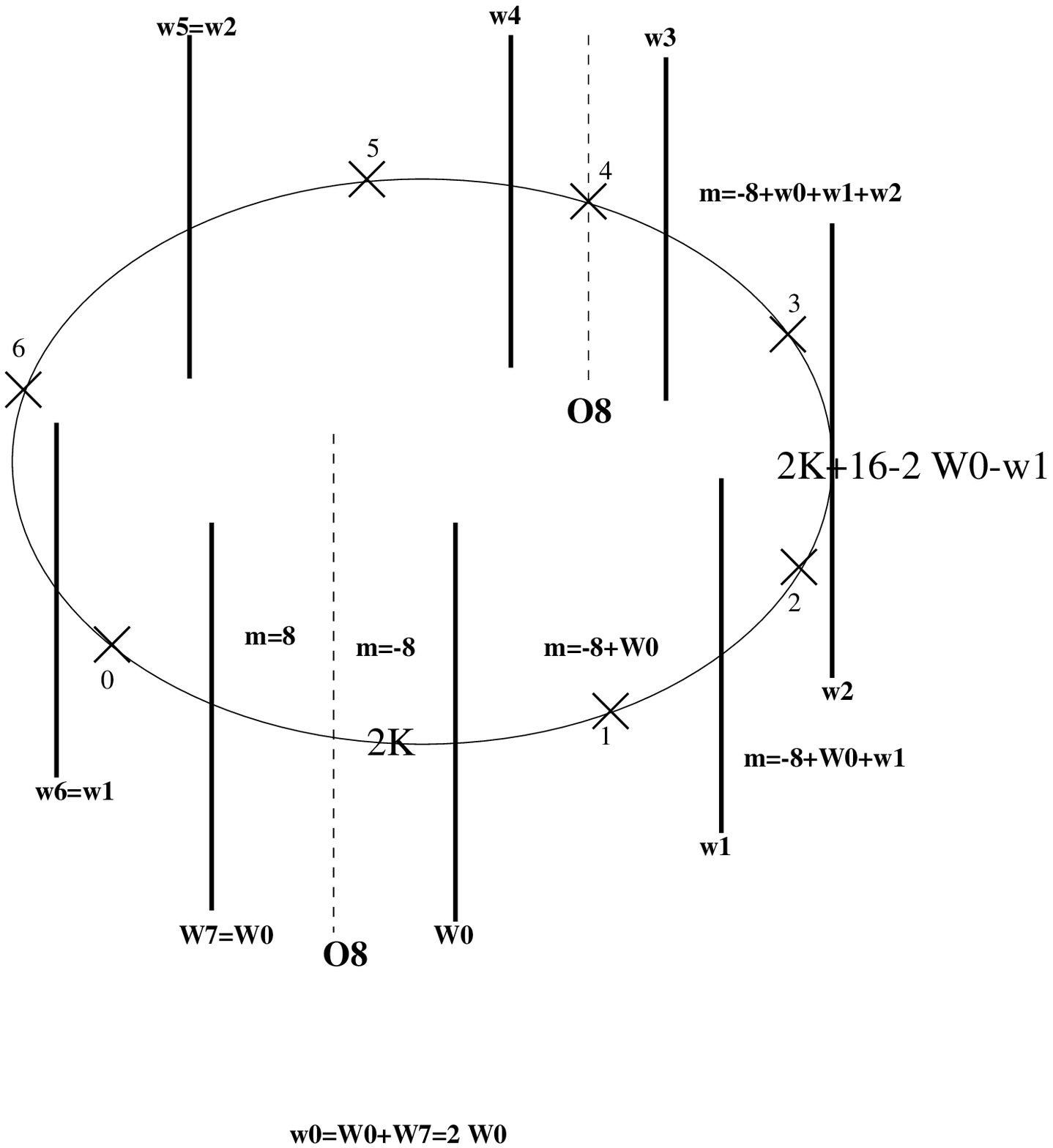}{10truecm}
\figlabel{\su3}
\vspace{1cm}
Figure \su3 illustrates the notation. Again it is useful to introduce
$W_{\mu}$s as defined in the previous section to take care of the effect
that for every gauge group whose color branes stretch over the
orientifold half of the $w_{\mu}$ branes have to be located on either
side. Therefore in these cases we define $W_{\mu} = \half w_{\mu}$ while
in all other cases $W_{\mu}= w_{\mu}$. These special values of $\mu$
are 0 and $(k+1)/2$ for even $k+1$ with vector structure, just $0$
for $k+1$ odd and there is no such gauge group for $k+1$ even
and without vector structure, as can be easily seen from the identification
in terms of brane positions given above.

\subsubsection*{$k+1$ odd}
We are studying $K$ 5 branes on an A$_k$ singularity on the IIB side. That
is we consider $k+1$ 5 branes on the circle with $K$ D6
branes stretching between them in the HW picture.
As in Figure \sp_cp we have two orientifolds on the circle, both with
negative charge. One of the 5 branes is stuck on the one orientifold.
We label the 5 branes with $\mu$ running from 0 to $k$ as in Figure \su3
\footnote{Note that in this numbering the stuck 5 brane is number $(k+2)/2$.}.
In addition there are 32 D8 branes. Let  again $w_{\mu}$ denote the number of
8 branes in between the $\mu$th and $(\mu +1)$th NS brane. Obviously
$\sum_{\mu} w_{\mu}=32$. Also symmetry with respect
to the orientifold plane requires $w_{\mu}=w_{k+1-\mu}$.
According
to the rules of how to introduce 8 branes and orientifolds
\footnote{The D8 branes have two effects: first they introduce a fundamental
hypermultiplet in the gauge group they are sitting in, second they decrease
the number of colors in every following gauge group
to their right by one per 5 brane in between,
since the value of the cosmological constant
$m$ changed.} , we find that the gauge group
is 
\begin{equation}
\label{odd2}
Sp(V_{0}) \times \prod_{\mu=1}^{k/2} SU(V_{\mu})
\end{equation}
with
$$ V_{\mu} = 2K - \sum_{\nu=0}^{\mu-1} (\mu-\nu) W_{\nu} + 8 \mu $$

We have
$\half w_0 \fund_0, \oplus_{\mu=1}^{k/2} w_{\mu} \fund_{\mu},
 \oplus_{\mu=1}^{k/2}
(\fund_{\mu-1}, \fund_{\mu})$ and
$\asym_{k/2}$ matter multiplets (subscripts label
the gauge group) and $k/2$ tensor multiplets.

Using the symmetry property $W_{\mu}=w_{\mu}=w_{k+1-\mu}=W_{k+1-\mu}$, 
$W_{0} = \half w_0$ the requirement of having a total
of 32 D8 branes $\sum_{\nu=0}^{k} w_{\nu} =32$ can be rewritten
as $$\sum_{\nu=0}^{k/2} W_{\nu} = 16 $$
which just says that 16 of the 32 D8 branes have to be on either
side of the orientifold.
Therefore $$\sum_{\nu=0}^{\mu-1} \mu W_{\nu} = 16 \mu -
 \mu \sum_{\nu=\mu}^{k/2} W_{\nu}$$
and (\ref{odd2}) becomes
$$V_{\mu} = 2K - 8 \mu + \sum_{\nu=0}^{\mu-1} \nu W_{\nu} +
\sum_{\nu=\mu}^{k/2} \mu W_{\nu} $$

This gauge group and matter content
agrees with (\ref{odd}). This shows that our identification of $w_{\mu}$
from above, characterizing the Wilson lines at infinity, and
the $w_{\mu}$ we just defined in terms of brane positions indeed is correct.
If one wants to study the fixed point associated to this theory, one
has to bring the 5 branes together on top of each other.
In order to do so one has to move 5 branes through 8 branes, whose position
is according to the general philosophy a parameter and not a modulus.
Note that the gauge group and matter content
is not changed by doing so if one is taking into account that whenever
a 5 brane crosses a D8 a new 6 brane is created stretching between
them according to the usual HW effect \cite{hw}. However our identification
of the $w_{\mu}$ with 8 brane positions should not be applied to this
situation anymore. 
Since the $k$th
5 brane is stuck, we move all others on top of this one.
In the orbifold language this corresponds to choosing
the $0$th gauge group factor to be the IR free one. Any other choice
would indeed lead to another fixed point. However the $k$th 5 brane
can not participate in this and so these other fixed points are
already included in the discussion of lower values of $k$.
$k=0$ reduces to $Sp(2K)$ gauge theory with 16 $\fund$ and 1 $\asym$
matter multiplet and no tensors, 
the theory on $K$ $SO(32)$ heterotic 5 branes in a non-singular
space. This is what we expect, since the 1 NS brane from the HW
picture becomes a single KK monopole in the IIB picture that
reduces to flat space in the decompactification limit.

\subsubsection*{$k+1$ even with vector structure}
If we have A$_k$ with an even number $k+1$ of 5 branes, we have two choices.
We can have all 5 branes free to move around on the circle of
Figure \sp_cp or we have one 5 brane stuck at each orientifold separately.
Here we will discuss the former case and by comparing the
resulting gauge theory see that it corresponds to the case with
vector structure. Again we have $w_{\mu}$ D8 branes between the
$\mu$th and $(\mu +1)$th NS5 brane. Similar to above we find
perfect agreement with the orbifold analysis (\ref{evenwith}).
With no 5 branes we have an $Sp(2K)$ gauge theory
with 16 $\fund$ and 1 $\asym$
matter multiplet and no tensors, which is again, as expected,
just the $SO(32)$ 5 brane.

\subsubsection*{$k+1$ even without vector structure}
As in the previous case we study $k+1$ 5 branes on the circle with
two orientifold planes and 32 D8 brane with $k+1$ even. This time
one 5 brane is stuck at every one of the 2 O8 planes. The others
are free to move around. We find
again perfect agreement with the orbifold analysis (\ref{evenout}).
The simplest case with just the two stuck 5 branes has $SU(2K)$
gauge theory with 2 $\asym$, 16 $\fund$ matter fields and no tensors.
The corresponding gauge theory has no non-trivial fixed point. However it 
defines a `little string theory', as pointed out in \cite{intstring}. 

\subsection{Theories with Twisted 5 Branes}

From the analysis of 4d Hanany-Witten setups it is known that there
are two distinct ways to introduce orthogonal or symplectic
gauge groups: either by including and O6 along the D6 flavor
branes or an O4 along the D4 color branes. The O6 becomes
the O8 after T-dualizing to 6 dimensions. This is the case we studied so
far. The other possibility is to include an O6 along the D6 branes
in our setup.

\vspace{1cm}
\fig{Brane configuration yielding an alternating product of
$SO(2K+8)$ and $Sp(2K)$ gauge groups with bifundamentals}
{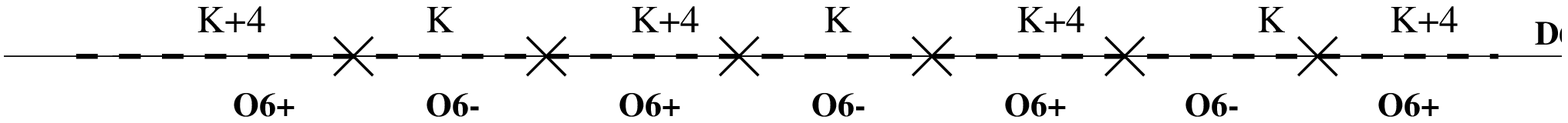}{14truecm}
\figlabel{\so_sp}
\vspace{1cm}

As explained in Section 2.4. this setup shown in Figure \so_sp
yields an
$$ \left \{ Sp(2K) \times SO(2K+8) \right \}^{(k+1)/2} $$
gauge group with bifundamentals and $k$ tensors. This can
hence be identified with the theories arising from the twisted 5 brane
(\ref{twist}).
The requirement that we have an even number of 5 branes is needed in
order to be able to compactify the circle (the gauge
group on the far right has to be the same as the one on the far left).
This is required
for the interpretation of our setup as a little string theory
and to be able to go to the dual picture. However from the HW
picture we also get consistent gauge theories with an odd number of 5 branes,
just yielding another $SO(2K+8)$ or $Sp(2K)$ factor. They 
probably lead
to new non-trivial fixed points.

 Now it easy to reformulate this theory
in terms of 5 branes at orbifolds. Under T-duality the O4 turns into
an IIB O5 plane. Under S-duality the O5 becomes an U5, like for
example discussed recently in \cite{6dwitten}. The U5 has exactly
the same properties with respect to NS5 branes like the O5 for D5 branes. 
It carries for example charge under the 6-form NS gauge field under
which the NS brane is charged. It enhances the (1,1) SUSY gauge theory
on the worldvolume of the NS5 brane to $SO$ or $Sp$ respectively.
The existence of such an object is required by S-duality. We thus
see that the twisted 5 brane of \cite{intblum} is just an NS5 brane
on top of the U5.

\section{New Fixed Points}

It is clear that there are several fixpoints that can not be realized via
the NS5, D6, D8 and O8 configuration we presented here. They involve all
the theories obtained by studying ALE spaces with D or E type singularity.
Our approach relied on the fact that upon T-duality in a transverse
direction IIB with an
A$_k$ type singularity turned into $k$ NS 5 branes. However no statement
like this is known for the D and E type singularities
\footnote{For IIA on a D-type singularity there is a dual IIB
description in terms of NS5 branes on an U5 orientifold plane, the
S-dual partner of the O5 \cite{6dwitten}}.
Therefore there are clearly many fixpoints that can only be realized in
the picture of branes at orbifolds.

However there are also some examples of theories that can only
be constructed in the Hanany-Witten type setup. Once we don't enforce
the $x_6$ to be compact we are basically free to
\begin{itemize}
\item put more 8 branes in the picture
\item use a different sign for the orientifold projection
\end{itemize}

If the circle is compact RR charge conservation forces us to have no
D8 branes in the case without orientifold and precisely 32 D8 branes
once we introduced the O8. In addition we are only allowed to
use the O8 with negative D-brane charge in order to be
able to cancel its charge with D8 branes. Once we are free
to have some surplus of RR charge we are free to use the
`other sign' for the orientifold projection, yielding an O8
with charge +16 and hence $SO$ groups or symmetric tensors instead
of $Sp$ groups and antisymmetric tensors like we had so far. In the
following we will present some of the theories that can be obtained
this way. It was already pointed out in \cite{intblum} that their
anomalies vanish. By embedding them in string theory we show that they
actually exist as fixpoint theories.
Since it was argued in \cite{consis} that 1 uncompact dimension is
not enough to let RR flux escape to infinity we should probably
limit ourselves to 32 D8 branes or one O8 with charge +16, so
that we can cancel the total charge by putting negatively charged O8
branes at infinity.

\subsection{Additional D8 Branes}

The simplest modification is to include additional 8 branes.
\vspace{1cm}
\fig{Product gauge group with additional 8 branes}
{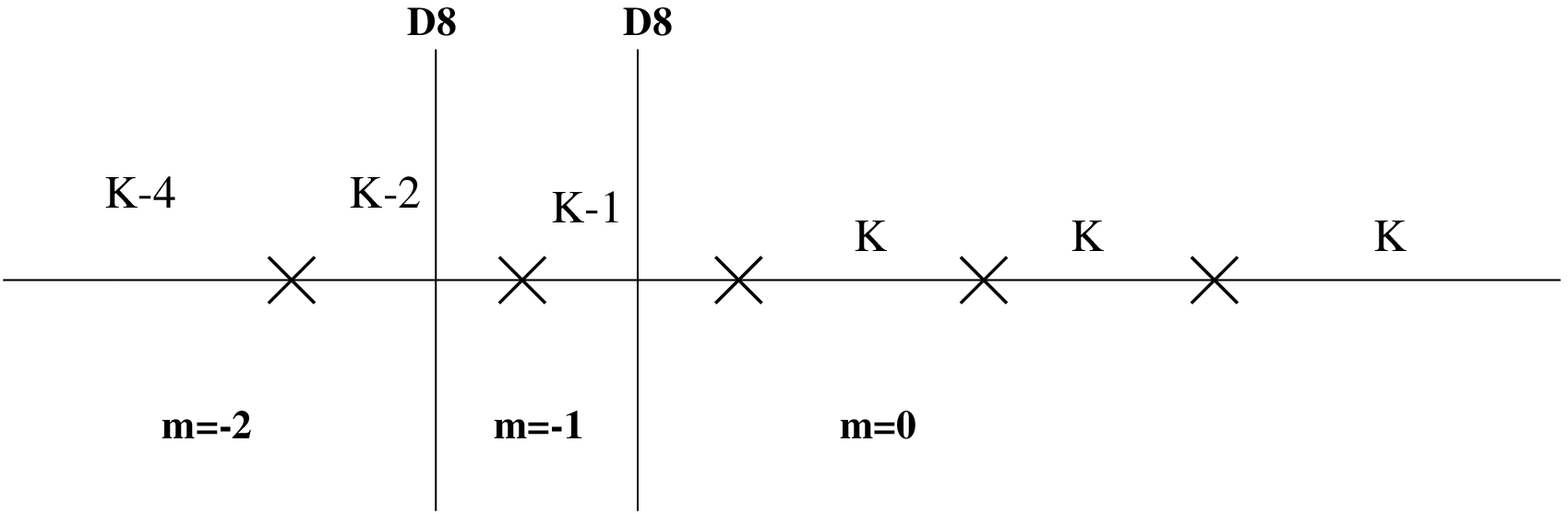}{15truecm}
\figlabel{\add8}
\vspace{1cm}

These theories are basically already obtained in the
analysis of section 4.2. We start out with the same gauge
group on the circle. Going to the fixed point means
that we move all the 5 branes together. In 4.2 we put them
all on top the O8. We might as well bring them together somewhere else.
\vspace{1cm}
\fig{Theory from Figure \protect \add8 arising from the
gauge theory of Figure \protect \sp_cp 
at a different point in moduli space}
{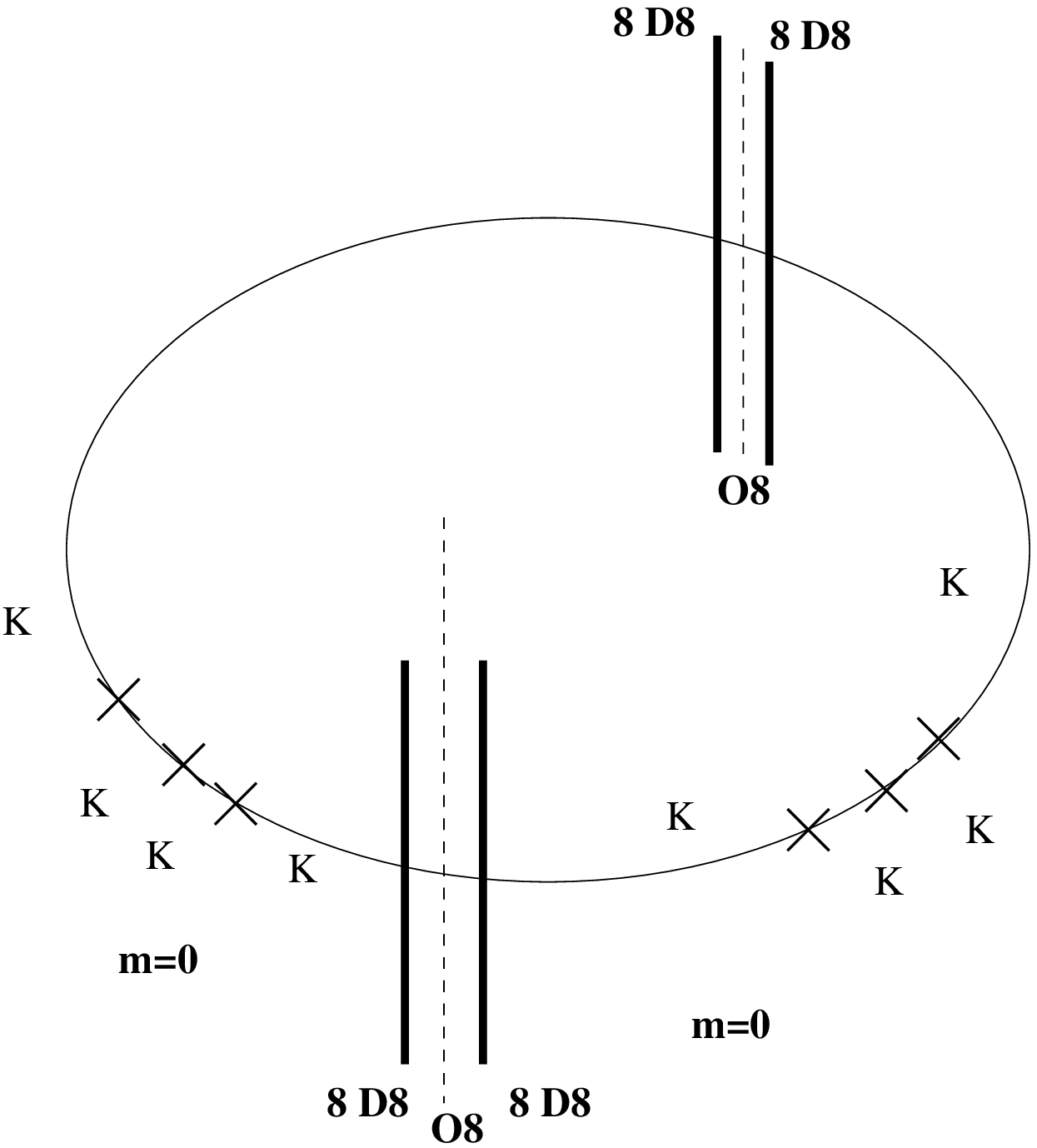}{9truecm}
\figlabel{\newsu}
\vspace{1cm}
Since the positions of the 5 branes correspond to the moduli from
the tensor multiplets this corresponds to a fixed point at a different
point in moduli space. Instead of decoupling one of the $Sp$ groups
we decoupled both of them. 

One interesting example of theories that can be constructed that way is
drawn in the following figure:
\vspace{1cm}
\fig{Gauge theory that also appears on an $E_8 \times E_8$
5 brane at an A$_{M-1}$ singularity}
{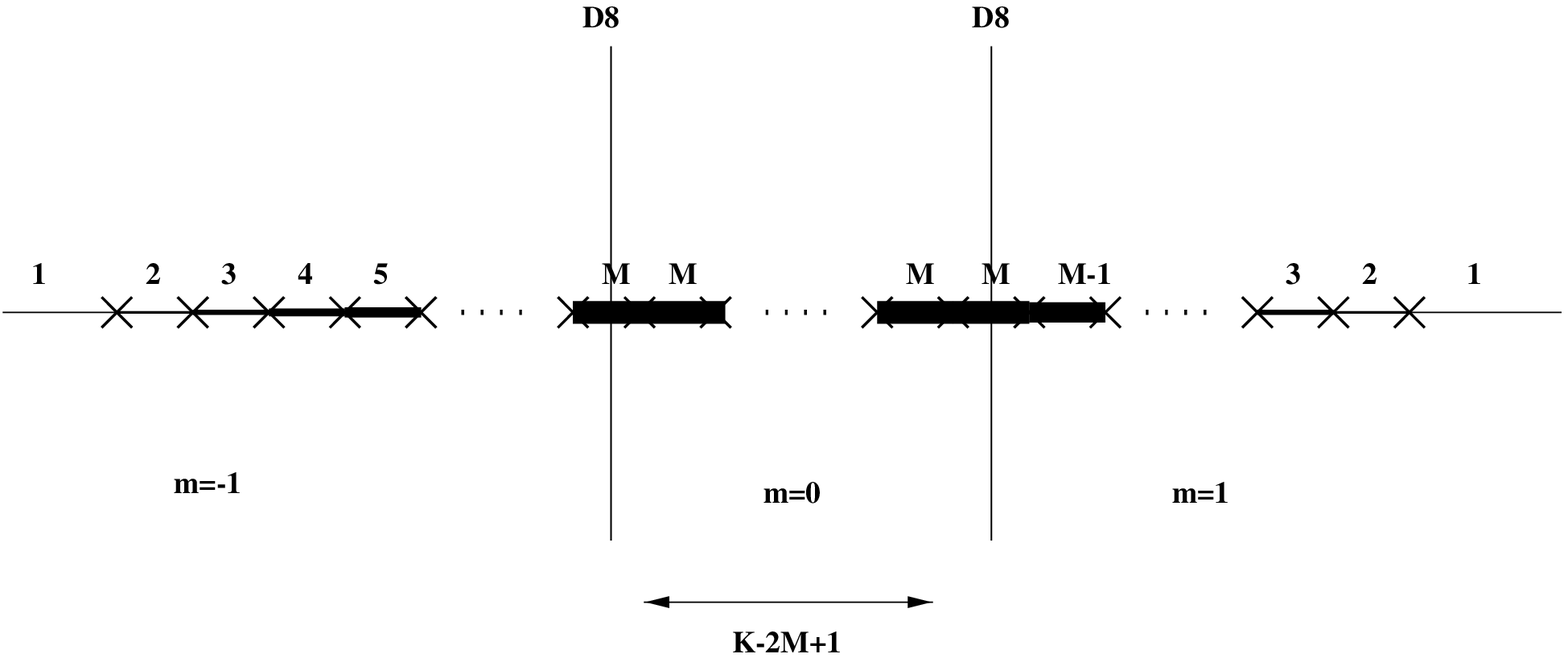}{15truecm}
\figlabel{\e8}
\vspace{1cm}
The strong coupling fixed point we obtain from this gauge theory by
banging together all the 5 branes can also be obtained
by studying an $E_8 \times E_8$ 5 brane at an A$_{M-1}$ singularity
\cite{intstring}.
This theory was also analyzed in \cite{aspinmor} from F-theory.
Unlike the cases discussed above involving $SO(32)$ heterotic
5 branes here we found no obvious
duality relation mapping the two descriptions of this
fixed points onto each other.

\subsection{$SO$ groups and Symmetric Tensors}

As mentioned above we can introduce orientifold O8 planes with
positive 8-brane charge in our picture. 
As mentioned above, we should probably
nevertheless insist on having total RR charge zero. This can
be done by putting a negatively charged O8 at infinity. This
kind of IIA orbifold theory with oppositely charged orientifold planes
was recently analyzed in \cite{wittenorbi}

This way we can generate theories with
$$ SO(2K) \times \prod_{\mu=1}^{(k-1)/2}
SU(2K-8\mu) \times \; Sp(2K-4(k+1))$$
gauge group and matter content 
$ \oplus_{\mu=1}^{(k+1)/2}
(\fund_{\mu-1}, \fund_{\mu})$. In addition there are $(k+1)/2$ 
tensor multiplets.

$$ SO(2K) \times \prod_{\mu=1}^{k/2} SU(2K -8 \mu)$$
gauge group with  $
 \oplus_{\mu=1}^{k/2}
(\fund_{\mu-1}, \fund_{\mu})$ and
$\asym_{k/2}$ matter multiplets
and $k/2$ tensor multiplets.

$$ Sp(2K) \times \prod_{\mu=1}^{k/2} SU(2K +8 \mu)$$
gauge group with  $
 \oplus_{\mu=1}^{k/2}
(\fund_{\mu-1}, \fund_{\mu})$ and
$\sym$ matter multiplets
and $k/2$ tensor multiplets.

$$\prod_{\mu=0}^{(k-1)/2} SU(2K+8) $$
with $ 
 \oplus_{\mu=0}^{(k-1)/2}
(\fund_{\mu-1}, \fund_{\mu})$, $ \asym_{0}$ and
$\sym_{(k-1)/2}$ matter multiplets
and $(k-1)/2$ tensor multiplets.

These possibilities correspond to an even 
number of 5 branes free to move on the
circle (`with vector structure'), an odd number of branes with one
brane stuck on the positively or negatively charged orientifold
or an even number of branes with one stuck at each orientifold
(`without vector structure').

In these theories the $F^4$ anomaly vanishes and the $F^2$ part
is cancelled by the coupling to the tensors as above
\cite{intblum,sweden} . By embedding them
into string theory we showed that their corresponding strong
coupling fixed points exist.

Witten \cite{wittenorbi,parkes} presented a IIB dual description
of this orientifold. There the worldsheet parity
operation is combined with a $\pi$ shift on a compact spacetime
circle. In the same spirit we can apply
the T-duality used in this paper and thereby find that 
the fixed points of this section should
have a dual realization as the worldvolume theory of NS5 branes in 
orientifolded IIB on an ALF space, where the worldsheet parity
operation is combined with a $\pi$ shift on the compact ALF circle.

There are three cases when there are no tensor multiplets left
in the theory: just one 5 brane stuck to either one of the
oppositely charge orientifolds or one stuck to each. The resulting
gauge theories are
\begin{itemize}
\item $SO(2K)$ with an adjoint
\item $Sp(2K)$ with an adjoint
\item $SU(2K)$ with an symmetric and an antisymmetric tensor
\end{itemize}
In all three cases the anomaly vanishes exactly. Even though
these theories do not constitute new fixed points they nevertheless
define new `little string theories'.

\section*{Acknowledgements}
We would like to thank K. Behrndt, J. Distler, A. Hanany and D. L\"ust
for useful discussions. The work of
both of us is supported by the DFG.

\end{document}